# Adaptive driver-automation shared steering control via forearm surface electromyography measurement

Zheng Wang, *Member, IEEE,* Satoshi Suga, Edric John Cruz Nacpil, Zhanhong Yan, and Kimihiko Nakano, *Member, IEEE*

*Abstract*— Shared steering control has been developed to reduce driver workload while keeping the driver in the control loop. A driver could integrate visual sensory information from the road ahead and haptic sensory information from the steering wheel to achieve better driving performance. Previous studies suggest that, compared with adaptive automation authority, fixed automation authority is not always appropriate with respect to human factors. This paper focuses on designing an adaptive shared steering control system via sEMG (surface electromyography) measurement from the forearm of the driver, and evaluates the effect of the system on driver behavior during a double lane change task. The shared steering control was achieved through a haptic guidance system which provided active assistance torque on the steering wheel. Ten subjects participated in a high-fidelity driving simulator experiment. Two types of adaptive algorithms were investigated: haptic guidance decreases when driver grip strength increases (HG-Decrease), and haptic guidance increases when driver grip strength increases (HG-Increase). These two algorithms were compared to manual driving and two levels of fixed authority haptic guidance, for a total of five experimental conditions. Evaluation of the driving systems was based on two sets of dependent variables: objective measures of driver behavior and subjective measures of driver workload. The results indicate that the adaptive authority of HG-Decrease yielded lower driver workload and reduced the lane departure risk compared to manual driving and fixed authority haptic guidance.

## I. Introduction

DRIVING automation has drawn much attention in recent years, and remarkable outcomes have been attained with adaptive cruise control and lane keeping systems. However, before fully automated driving cars can be introduced into the market, keeping the driver in the driving control loop with a sense of agency is still essential [1, 2]. Shared steering control, which combines the abilities of human driver and the vehicle automation, has been developed as a suitable approach to keep the driver in the steering control loop, while reducing driver workload [3, 4].

Ideally, when driving with a shared steering control system, the driver can comfortably rely on haptic guidance torque to drive more safely. It has been found that haptic guidance can assist drivers with curve negotiation by producing proper direction and torque on the steering wheel [5]. When drivers have low attention caused by fatigued driving, the haptic guidance steering reduces their lane departure risk [6]. When visual information from the road ahead is degraded, as in the case of dense fog, haptic guidance steering yields better lane-following performance than manual driving [7, 8]. A haptic guidance system has also been designed to assist drivers with prompt and steady steering for emergency obstacle avoidance [9]. On the other hand, it is suggested that fixed automation authority is not always appropriate with regard to human factors, in comparison to adaptive automation authority [7]. One way to adjust the level of adaptive authority is to address environmental factors, including vehicle position, yaw rate, etc. [10, 11]. Another way is to adjust the authority based on the driver states, which has been drawn little attention.

This work was supported by a Grant-in-Aid for Early-Career Scientists (No. 19K20318) from the Japan Society for the Promotion of Science.

Z. Wang is with the Institute of Industrial Science, The University of Tokyo, Tokyo 153-8505, Japan (corresponding author: 81-03-5452-6916, e-mail: z-wang@iis.u-tokyo.ac.jp).

S. Suga, is with the Institute of Industrial Science, The University of Tokyo, Tokyo 153-8505, Japan; Technical University of Darmstadt, Germany, Karolinenplatz 5, 64289 Darmstadt, Germany (e-mail: satosuga95@gmail.com)

E. Nacpil is with the Institute of Industrial Science, The University of Tokyo, Tokyo 153-8505, Japan (e-mail: enacpil@iis.u-tokyo.ac.jp)

Z. Yan is with the Institute of Industrial Science, The University of Tokyo, Tokyo 153-8505, Japan (e-mail: yanzhjob@iis.u-tokyo.ac.jp)

K. Nakano is with the Institute of Industrial Science, The University of Tokyo, Tokyo 153-8505, Japan (e-mail: knakano@iis.u-tokyo.ac.jp).

Haptic guidance has not only been applied for driver steering assistance, but also for human-robot cooperation, including tele-operation [12, 13], exoskeleton control [14, 15], and upper-limb rehabilitation [16]. Human-robot physical interaction in such cases was achieved through providing haptic feedback, and human muscle activity was measured as sEMG signals to adjust control authority of automation, which resulted in better human-robot cooperation [13, 14].

When driving with a haptic guidance system, drivers could integrate visual sensory information from road ahead and haptic sensory information from steering wheel to achieve better driving performance [17]. It has been found that the drivers could choose to be relaxed or to resist the haptic guidance by adjusting their arm admittance [3]. Grip force and sEMG signal have been measured to estimate the muscle fatigue while holding steering wheel in different positions [18]. Moreover, previous research has suggested the relationship between arm admittance and grip strength on the steering wheel [19], which can be measured by sEMG signals from forearm muscles [20, 21].

Currently, how the authority of shared steering control should be adjusted with forearm muscle activity is unresolved. The aim of this study is to design an adaptive shared steering control system via forearm sEMG measurement, and to evaluate its effect on driving performance and driver workload during a double lane change task. A schematic diagram of the shared steering control system is shown in Fig. 1. One hypothesis of this study is that haptic guidance with adaptive authority would yield lower driver workload and reduce the risk of lane departure compared to fix authority haptic guidance and manual driving. It is also hypothesized that driver grip strength would be different among the driving conditions due to mutual adaptation between the driver and the haptic guidance system.

This paper is organized as follows. Section II describes an experiment conducted with a high-fidelity driving simulator, including details of the participants, apparatus, scenario, and evaluation method. Section III illustrates the experimental results, followed by Section IV in which the effects of different types haptic guidance on driver behavior are discussed. Finally, the conclusion is conveyed in Section V.

## II. Experimental Method

### A. Participants

Ten healthy male subjects were recruited to participate in the experiment. Their age ranged from 20 to 25 (mean = 22, SD = 1.8). All had a valid Japanese driver's license with a driving experience (mean = 2.6 years, SD = 1.3). In response to the question of driving frequency, two participants reported driving once a week, and others less than once a week. They all had normal or corrected-to-normal vision when performing the driving tasks in the experiment.

The experiment was approved by the Office for Life Science Research Ethics and Safety, Graduate School of Interdisciplinary Information Studies, the University of Tokyo (No. 14 in 2017). Each participant received monetary compensation for the involvement in the experiment.

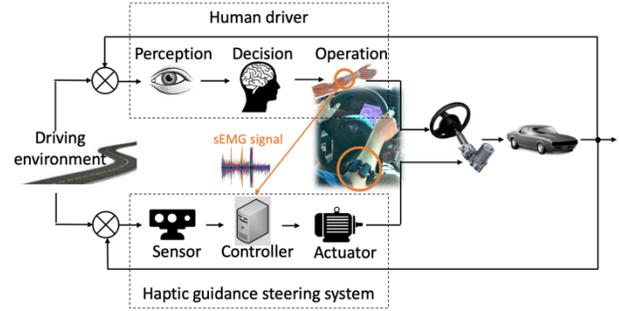

Fig. 1. Schematic diagram of shared steering control.

### B. Apparatus

As shown in Fig. 2, the experiment was conducted in a moving-based driving simulator with brake and accelerator pedals, an actuated steering wheel, and an instrument dashboard. The driving simulator is considered to be high-fidelity because it includes a 140-degree field-of-view and a moving platform with six degrees of freedom.

To emulate the feeling of on-road driving, high frequency vibrations were produced by the moving platform, engine sounds were provided by two stereo speakers, and a self-aligning torque was generated by the actuated steering wheel. Raw data of driving performance were recorded in the host computer of the driving simulator at a sample rate of 120 Hz.

A Myo armband system (Thalmic Labs, Inc.) was employed to measure the sEMG signal of driver's dominant forearm. Compared to other commercially available armbands, Myo armband provides software development kit which allows developers to obtain access to measured data. It has been widely used in research experiments and applications [22, 23]. The muscle activity was measured by calculating the root mean square (RMS) value of the activation from eight sEMG sensors within the armband [24]. Normalization of driver grip strength was performed by measuring the maximum sEMG value for each participant before the formal experiment. In the experimental conditions of adaptive authority haptic guidance, the authority of haptic guidance is correlated to the normalized sEMG value.

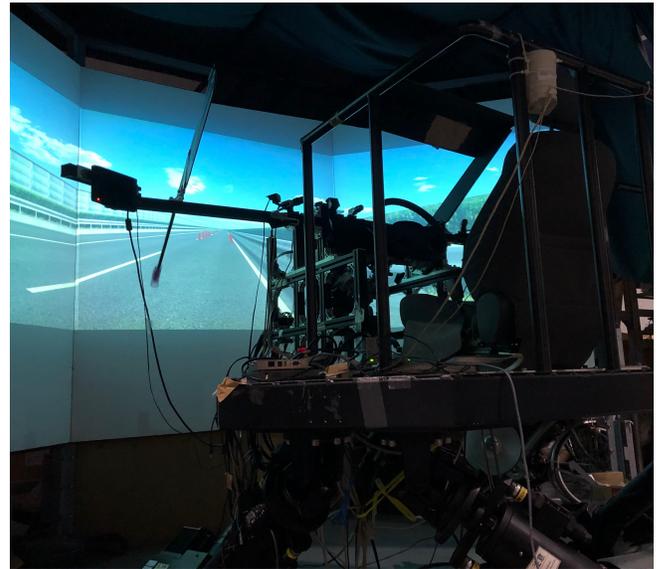

Fig. 2. High-fidelity driving simulator used in the experiment.

### C. Haptic Guidance Steering System

In the driving simulator, an electronic steering system was connected to the host computer through a CAN (controller area network) communication. The electronic steering system mainly consisted of a steering wheel, a servo motor and an electronic control unit (ECU), as shown in Fig. 3. The variables shown in Fig. 3 are defined in Table I. After calculating the real-time haptic guidance torque, the host computer inputted it to the ECU to actuate the servomotor that applied haptic guidance torque to the steering wheel. The maximum motor torque was 5 N·m, and the motor reduction ratio was 1/14. Accordingly, the steering wheel system can provide 70 N·m as a maximum active torque. The sensor resolution was 0.1° for the steering wheel angle, and 0.005 N·m for the steering torque.

The magnitude and direction of haptic guidance torque were determined by comparing the vehicle trajectory and target trajectory. It is important to notice that the system is designed to assist the driver, instead of controlling the vehicle automatically. Consequently, the driver can choose to overrule the system at any time by exerting more torque on the steering wheel. In this experiment, the haptic guidance torque was limited to 5 N·m [6].

Parameters of the vehicle trajectory and target trajectory are presented in Fig. 4. The target trajectory is generated by a 5$^{th}$

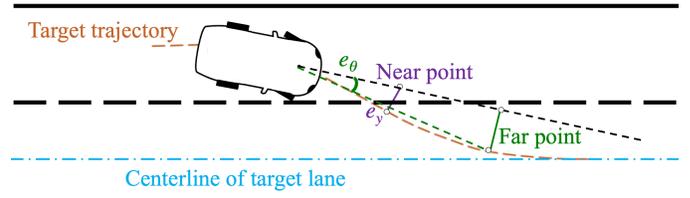

Fig. 4. Diagram of two-point haptic guidance model following target trajectory.

degree Bezier curve that was previously reported to achieve smooth lane changing performance [25, 26]. The coordinates of the target trajectory were stored in the host computer of the driving simulator. The haptic guidance torque was calculated based on a model with two look-ahead points [27]. The lateral error at the near point, $e_{y(\text{near})}$, is defined as the distance between the position of vehicle and target trajectory at the near point, and the yaw error at the far point, $e_{\theta(\text{far})}$, is defined as the angle between the direction of vehicle and target trajectory at the far point. The look-ahead time of the near point is 0.3 s, and the far point is 0.7 s.

The haptic guidance torque, $T_h$, is expressed as

$$T_h = K(a_1 e_{y(\text{near})} + a_2 e_{\theta(\text{far})}) \quad (1)$$

where $a_1$ and $a_2$ are constant gains for $e_{y(near)}$, and $e_{\theta(far)}$, respectively; $K$ is constant gain for the haptic guidance torque. The gains of $a_1$ and $a_2$ were determined as 0.19 and 3.8 respectively by a trial-and-error process.

The settings of shared steering control authority were inspired by [12], in which different types of fixed and adaptive haptic guidance were addressed. In this experiment, different types of haptic guidance were determined by the value of $K$, as shown in Table II: The haptic guidance with a strong feedback gain (HG-Strong, $K = 1.0$) was half of the gain set for automated double lane change; the haptic guidance with a normal feedback gain (HG-Normal, $K = 0.5$) was half of the gain set for HG-Strong; in the condition of HG-Decrease, the gain of haptic guidance decreases linearly from 1 to 0, when the grip strength increases from 0 to the maximum value, $sEMG_{REF}$; and in the condition of HG-Increase, the gain of haptic guidance increases linearly from 0 to 1, when the grip strength increases from 0 to the maximum value, $sEMG_{REF}$. Moreover, if the grip strength exceeded $sEMG_{REF}$ during the double lane change task, the gain of haptic guidance gain was set to 0 for HG-Decrease, and 1.0 for HG-Increase.

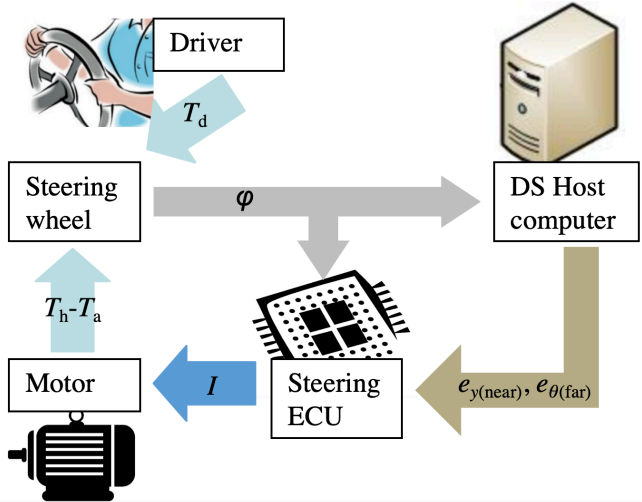

Fig. 3. Diagram of steering control loop.

TABLE I
SYMBOLS IN THE STEERING CONTROL LOOP

| | Definition |
|---|---|
| $e_{y(\text{near})}$ | Lateral error at the near point |
| $e_{\theta(\text{far})}$ | Yaw error at the far point |
| $I$ | Motor control current |
| $T_d$ | Driver input torque |
| $T_h$ | Haptic guidance torque |
| $T_a$ | Aligning torque |
| $\varphi$ | Steering wheel angle |

TABLE II
EXPERIMENTAL CONDITIONS WITH DIFFERENT TYPES OF HAPTIC GUIDANCE

| Condition | Description | Feedback gain ($K$) |
|---|---|---|
| Manual | No haptic guidance | 0 |
| HG-Strong | Fix authority with strong feedback | 1.0 |
| HG-Normal | Fix authority with normal feedback | 0.5 |
| HG-Decrease | Adaptive authority with decreased feedback as grip strength increases | $1.0 - \frac{sEMG}{sEMG_{REF}}$ |
| HG-Increase | Adaptive authority with increased feedback as grip strength increases | $\frac{sEMG}{sEMG_{REF}}$ |

## D. Experimental Conditions and Scenario

The participants drove under five conditions, as shown in Table II, each with a different type of haptic guidance: (1) No haptic guidance (Manual), (2) haptic guidance with a strong feedback gain (HG-Strong), (3) haptic guidance with a normal feedback gain (HG-Normal), (4) haptic guidance decreases when grip strength increases (HG-Decrease), and (5) haptic guidance increases when grip strength increases (HG-Increase).

Two Latin squares were used to partially counterbalance the within-subject order of the conditions among the 10 participants. The first Latin square with all five experimental conditions was used for the first through fifth participant, whereas the second Latin square for the sixth through tenth participant mirrored first Latin Square.

As shown in Fig. 5, the driving environment was a two-lane expressway with lanes 3.6 m wide and an emergency lane on the left. Lane markings were solid lines and dashed lines. An overhead view of the double lane change track with cones is shown in Fig. 6.

The driving speed of the ego vehicle was fixed at 50 km/h by a PID controller; thus, the participants did not need to operate the accelerator and brake pedals. Since steering and speed are interdependent, the speed was fixed to prevent it from confounding the assessment of steering performance.

## E. Procedure

Prior to participating in the experiment, test subjects signed a consent form explaining the procedure of the experiment. Each participant mounted the Myo armband on the dominant forearm. Experiment preparation consisted of two procedures: (1) Calibration and (2) normalization.

In order to calibrate the Myo armband for each participant, the default program supplied by the armband manufacturer was used. Normalization of grip strength was realized by having each participant grip the driving simulator steering wheel in a "ten-to-two" position for two seconds with maximum effort. This procedure was repeated three times with 10 s of rest between each repetition. The mean value across all repetitions for a given participant was used as the reference sEMG value, $sEMG_{REF}$, for normalization.

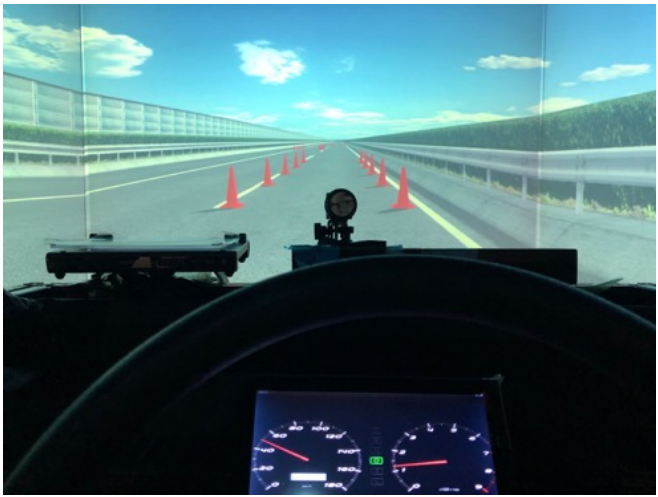

Fig. 5. Driving environment in the experiment.

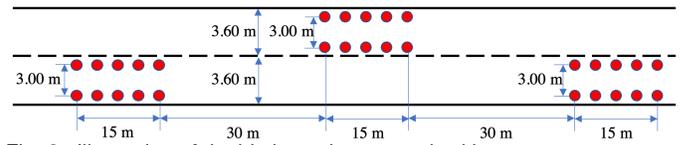

Fig. 6. Illustration of double lane change track with cones.

After the preparation, the participants boarded the driving simulator, and were instructed to grab the steering wheel in the "ten-to-two" position and to follow Japanese traffic regulations. Before the experimental session, the participants were required to perform a practice task to become familiar with the driving simulator. In the formal experimental session, the participants performed the double lane change task five times under each of the five driving conditions. Therefore, there were 25 trials of double lane change per participant. After each condition, the participants were asked to complete a questionnaire to assess subjective task load. The entire experiment took approximately 90 min per participant.

## F. Measures

The double lane change performance was measured by driver steering behavior, lane departure risk, sEMG signal, and subjective assessment on driver workload.

Driver steering behavior was evaluated by measuring driver input torque and steering wheel angle. RMS value of driver input torque was calculated to evaluate driver steering effort. The steering control activity was evaluated by calculating the RMS value of steering wheel angle and steering wheel peak angle. The steering wheel peak angle includes maximum positive steering wheel angle and minimum negative steering wheel angle.

Lane departure risk during the double lane change was evaluated by calculating lateral error with respect to the centerline of the lane. The double lane change process consists of two parts: first lane change part and second lane change part. At the first lane change ending point when the vehicle drives parallel with respect to the entered lane, the lateral error to centerline of lane was calculated. At the second lane change ending point, the driver normally made a small adjustment to get back to the centerline of lane which is called overshoot, and the lateral error to centerline of lane was calculated.

The sEMG signals of forearm muscle were measured to examine driver grip strength during the double lane change task. The RMS value of normalized sEMG signal ($sEMG/sEMG_{REF}$) was calculated, and a higher value would indicate a larger grip strength.

The perceived workload of driving task was rated using the NASA-TLX [28]. The participants were asked to use the NASA-TLX to assess their workload at the end of each driving condition. In the first step, each item of the index was investigated separately to obtain the scale score. In the second step, an individual weighting of the items was explored to obtain the weighted score. The weighted score was multiplied by the scale score for each item, and then the overall task load score was obtained.

## G. Data Analysis

Data were statistically analyzed using one-way repeated measures ANOVA with the Fisher-Hayter Post Hoc test to determine whether there was any significant difference between

TABLE III
MEANS AND STANDARD DEVIATIONS OF THE DEPENDENT MEASURES OF DRIVER BEHAVIOR

| Variable | Manual (1) M(SD) | HG-Strong (2) M(SD) | HG-Normal (3) M(SD) | HG-Dec (4) M(SD) | HG-Inc (5) M(SD) | p | 1-2 | 1-3 | 1-4 | 1-5 | 2-3 | 2-4 | 2-5 | 3-4 | 3-5 | 4-5 |
|---|---|---|---|---|---|---|---|---|---|---|---|---|---|---|---|---|
| RMS of driver input torque (N·m) | 1.096 (0.072) | 0.596 (0.140) | 0.786 (0.092) | 0.657 (0.149) | 0.884 (0.142) | < 0.001 | *** | *** | *** | *** | *** | 0.49 | *** | * | 0.12 | * |
| RMS of SWA (deg) | 24.1 (6.8) | 26.5 (8.1) | 25.2 (6.8) | 23.2 (5.8) | 24.1 (6.2) | 0.009 | + | 0.29 | 0.64 | 1.00 | 0.71 | + | 0.16 | ** | 0.48 | 0.46 |
| Maximum positive value of SWA (deg) | 39.6 (16.2) | 47.2 (20.0) | 41.8 (17.8) | 38.3 (15.9) | 40.3 (15.7) | 0.002 | *** | 0.66 | 0.76 | 0.98 | 0.27 | ** | 0.11 | 0.19 | 0.89 | 0.20 |
| Minimum negative value of SWA (deg) | -44.5 (14.4) | -47.2 (16.9) | -44.1 (12.8) | -41.0 (12.7) | -44.4 (12.3) | 0.032 | 0.47 | 0.99 | *** | 1.00 | 0.64 | * | 0.70 | 0.11 | 1.00 | * |
| Lateral error at the end of 1st LC (m) | 0.439 (0.125) | 0.335 (0.203) | 0.397 (0.156) | 0.280 (0.165) | 0.406 (0.133) | 0.014 | + | 0.62 | * | 0.85 | 0.51 | 0.86 | 0.53 | + | 0.97 | * |
| Lateral error at the end of 2nd LC (m) | 0.211 (0.098) | 0.194 (0.122) | 0.229 (0.094) | 0.167 (0.073) | 0.241 (0.119) | 0.243 | 0.95 | 0.89 | 0.60 | 0.91 | 0.80 | 0.82 | 0.64 | 0.23 | 0.99 | * |
| RMS of normalized sEMG (%) | 8.10 (3.62) | 8.66 (4.18) | 7.86 (3.52) | 7.58 (3.72) | 7.86 (3.39) | 0.265 | 0.82 | 0.95 | 0.58 | 0.69 | 0.62 | 0.27 | 0.45 | 0.78 | 0.80 | 0.91 |

+p < 0.1, *p < 0.05, **p < 0.01, ***p < 0.001
SWA: Steering wheel angle; LC: Lane change.

the driving conditions. The significant level was set to $p < 0.05$ to reject null hypothesis that there was no significant difference. When a *p*-value is between 0.05 and 0.1, it is interpreted as a tendency towards significant difference.

## III. RESULTS

In this section, results are presented separately for driver steering behavior, lane departure risk, sEMG measurement and subjective assessment of driver workload. The results indicate that driver behavior was significantly influenced by different types of haptic guidance. Table III lists the means and standard deviations, results of one-way repeated measures ANOVA, and pairwise comparisons based on the measurements of driver behavior.

### A. Driver Steering Behavior

The results of driver steering behavior were obtained by measuring driver input torque and steering wheel angle.

As indicated in Table III, the RMS of driver input torque was significantly different among the five conditions. Fig. 7 shows the result of RMS of driver input torque with error bar among 10 participants. Driver input torque was significantly higher in the condition of Manual than other conditions, which indicates that haptic guidance significantly reduced driver steering effort. HG-Strong induced significantly lower driver input torque compared to other conditions, and HG-Decrease was in between of HG-Strong and HG-Normal. Moreover, HG-Decrease induced lower driver input torque compared to HG-Increase.

As indicated in Table III, RMS of steering wheel angle was significantly different among the five conditions. Fig. 8 shows the results of RMS of steering wheel angle with error bars among 10 participants. It can be observed that HG-Decrease yielded lower steering wheel angle compared to HG-Normal and HG-Strong, and Manual yielded lower steering wheel angle compared to HG-Strong. Therefore, driver steering control activity was to some extent reduced by HG-Decrease and increased by HG-Strong.

The result of steering wheel angle is also indicated by plotting the averaged steering wheel angle with standard deviation for experimental sessions, as shown in Fig. 9. It can be observed that HG-Strong induced relatively larger steering wheel angle and HG-decrease induced relatively smaller steering wheel angle, which is in accordance with the results of RMS of steering wheel angle. This tendency is especially more evident when looking at the peak steering wheel angle during the double lane change process. As shown in Table III, the maximum positive value of steering wheel angle was significantly higher for HG-Strong than for Manual and HG-Decrease. The minimum negative value of steering wheel angle was significantly lower for HG-Decrease than for Manual, HG-Strong, and HG-Increase.

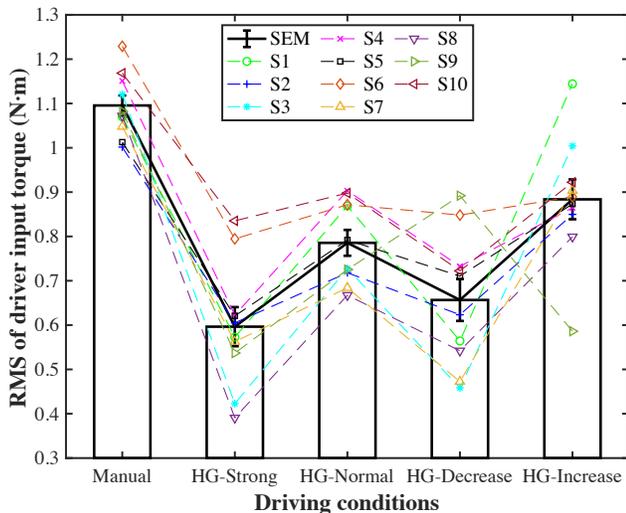

Fig. 7. RMS of driver input torque. Data error bars represent mean +/- SEM (standard error of mean).

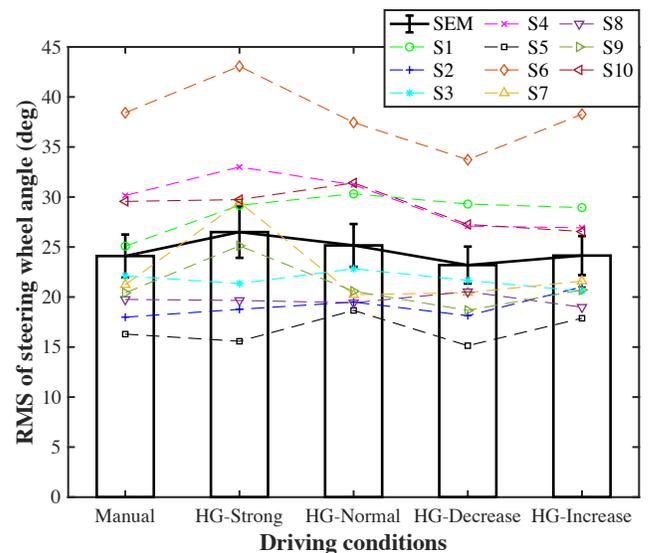

Fig. 8. RMS of steering wheel angle. Data error bars represent mean +/- SEM.

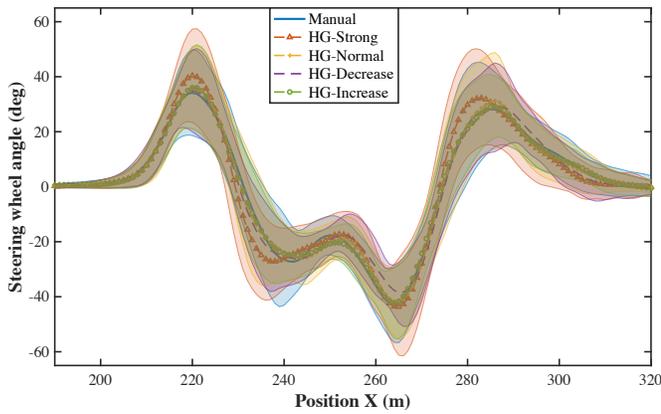

Fig. 9. Averaged steering wheel angle with standard deviation among five double lane change of all participants.

### B. Lane Departure Risk

The lane departure risk during the double lane change was evaluated by respectively calculating the lateral error from centerline of lane at the ending points of the first lane change part and second lane change part.

As indicated in Table III, lateral error at the end of first lane change part was significantly different among the five conditions. Fig. 10 shows the lateral error from centerline of lane with error bars among 10 participants. The lateral error was lower for HG-Decrease than for Manual, HG-Normal and HG-Increase, and lower for HG-Strong than for Manual. Therefore, HG-Decrease and HG-Strong reduced the lane departure risk compared to Manual.

As indicated in Table III, lateral error at the end of second lane change part was not significantly different among the five conditions. According to pairwise comparisons, the lateral error in the condition of HG-Decrease was lower than of HG-Increase. Fig. 11 shows the overshoot from centerline of lane with error bars among 10 participants. There is a tendency that HG-Decrease yielded a relatively lower lane departure risk compared to other conditions.

### C. sEMG Measurement

As indicated in Table III, RMS of normalized sEMG was not significantly different among the five conditions. Fig. 12 shows the results of RMS of normalized sEMG with error bars among 10 participants. There is a tendency that participants used more grip strength in the condition of HG-strong compared to other conditions, although it is not significant. It can be explained by that the participants tended to increase grip strength when driver-automation conflict occurred, and HG-Strong provided significantly larger active torques compared to the other conditions, as shown in Fig. 7.

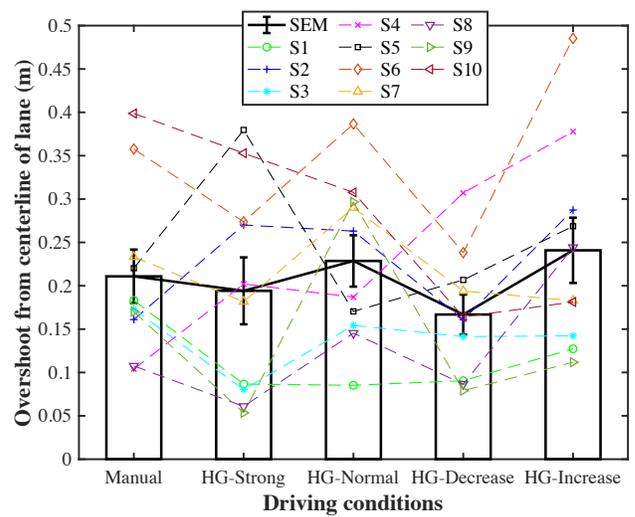

Fig. 11. Overshoot (lateral error) from centerline of lane at the end of second lane change part. Data error bars represent mean +/- SEM.

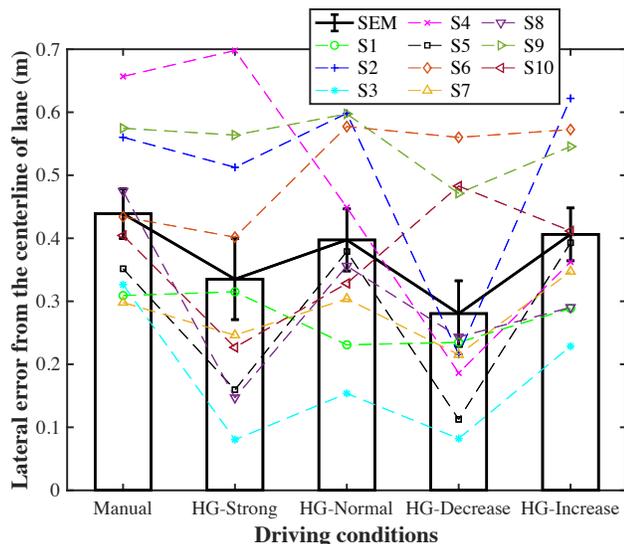

Fig. 10. Lateral error from centerline of lane at the end of first lane change part. Data error bars represent mean +/- SEM.

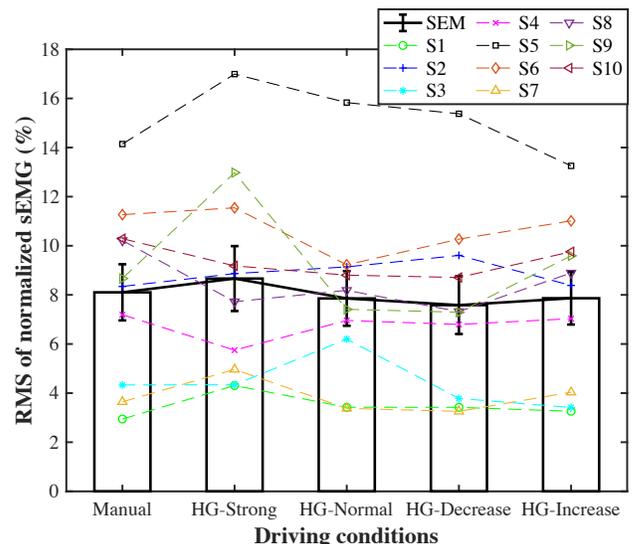

Fig. 12. RMS of normalized sEMG. Data error bars represent mean +/- SEM.

*D. Subjective Assessment of Driver Workload*

Subjective assessment of driver workload was rated by using NASA-TLX, and the results with error bars are shown in Fig. 13. Based on One-way repeated measures ANOVA, the statistical analysis shows that there was significant difference in Effort among the five conditions, where $p = 0.042$, and no significant difference for other parameters (Mental Demand: $p = 0.215$, Physical Demand: $p = 0.236$, Temporal Demand: $p = 0.341$, Performance: $p = 0.799$, Frustration: $p = 0.154$, Overall Workload: $p = 0.220$). The pairwise comparison results show that Effort was significantly lower for HG-Decrease than for Manual ($p < 0.05$), and significantly lower for HG-Increase than for Manual ($p < 0.01$). In addition, HG-Decrease ($p = 0.12$) and HG-Increase ($p = 0.12$) tended to reduce the overall workload compared to Manual, although the difference was not significant.

As shown in Fig. 13, there is a tendency that the driver workload could be reduced by haptic guidance compared to manual driving. As for the condition of HG-Decrease, the mean value shown in the figure is lower compared to the other conditions in terms of Mental Demand, Physical Demand, Effort, Frustration, and Overall Workload. Thus, HG-Decrease is to some extent more effective at reducing driver workload.

## IV. Discussion

Different types of driver–automation shared control were realized by using a haptic guidance steering system that was configured for fixed authority (HG-Strong and HG-Normal) and adaptive authority (HG-Decrease and HG-Increase). By looking at the results of driver steering behavior, both driver input torque and steering wheel angle were relatively lower for the condition of HG-Decrease, indicating that HG-Decrease yielded a lower steering control effort and control activity. This result is in accordance with the result of subjective assessment on driver workload by NASA-TLX, in which the driver workload was found to be lower for the condition of HG-Decrease. It is interesting that HG-Strong significantly reduced the driver input torque but increased the steering wheel angle. In some of the previous studies, only one of these two evaluation indexes was used to assess driver steering workload [8, 27]. HG-Strong shows controversial results in terms of these two indexes, indicating that both indexes should be addressed when analyzing driver steering workload.

It was expected that haptic guidance system would reduce lane departure risk relative to manual driving, and HG-Decrease would perform better relative to other haptic guidance conditions. The idea of designing HG-Decrease is to reduce the authority of haptic guidance when the driver increases the grip strength to gain more control authority. Although it was observed that HG-Decrease yielded a relatively lower lane departure risk compared to other conditions, other haptic guidance conditions did not have significantly better performance compared to manual driving. One explanation could be that the driver's individual driving trajectory [5] could be significantly different from the optimal $5^{th}$ degree Bezier curve that was used as target trajectory in the experiment. Another explanation could be that the driving speed of the ego vehicle was fixed at 50 km/h, whereas drivers would choose their own driving speed during regular driving [17]. If the driver could drive in a more regular way, and the target trajectory could be individualized, it is expected that the haptic guidance system could significantly reduce the lane departure risk.

One of the experimental hypotheses is that driver grip strength would be different among the driving conditions due to mutual adaptation between the driver and haptic guidance system. Human beings receive and integrate multiple sensory information from their surroundings, in order to understand and interact with the world. When driving with the haptic guidance system, drivers tended to integrate visual and haptic information to achieve better driving performance [8, 27]. It has been found that the driver could choose to rely on or to resist the haptic guidance by adjusting their arm admittance [3]. However, according to the sEMG measurement results, this expected outcome was not observed. One explanation could be that more time is needed for the driver to adapt to the haptic guidance system [27]. A longer driving scenario with curved roads will be designed as future work, and the mutual adaption behavior between driver and haptic guidance system will be investigated for lane change and lane following tasks.

Although the current experiment was conducted in a high-fidelity driving simulator, future work will address how driver behavior would be different in a real-vehicle experiment, especially for driver workload. Moreover, the subjective preference for different types of haptic guidance systems and whether the systems are preferable under realistic driving conditions could be tested in a future experiment with a real vehicle. Another limitation of the current experiment is that using the Myo armband instead of conventional sEMG equipment may limit the applicability of the study to other sEMG-based driving simulator experiments. The comparison between conventional and commercial sEMG equipment should be addressed by an experimental study in the future.

## V. Conclusion

This paper focuses on designing adaptive shared steering control via forearm sEMG measurement and evaluating its effect on driving performance and driver workload during a double lane change task. A driving simulator experiment was conducted, and the driving conditions were Manual, HG-Strong, HG-Normal, HG-Decrease, and HG-Increase.

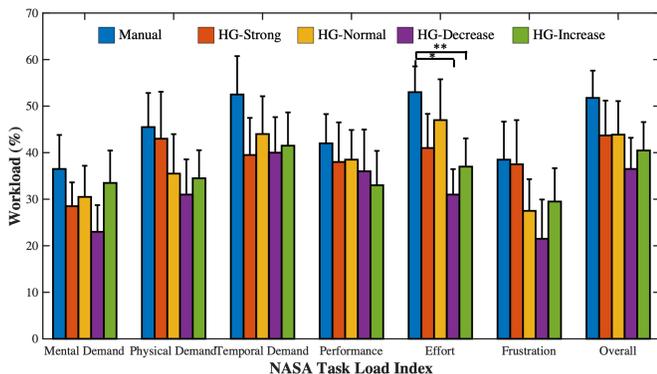

Fig. 13. Mean scores on NASA-TLX (*$p < 0.05$ and **$p < 0.01$). Data error bars represent mean + SEM.

The results show that the driver behavior, in terms of steering behavior, lane departure risk and driver workload, was different between driving with adaptive authority haptic guidance and with fixed authority haptic guidance. A reduction in both lane departure risk and driver workload was found in the condition of HG-Decrease compared to manual driving and fixed authority haptic guidance. This outcome suggests the potential of the adaptive authority of HG-Decrease to improve driver-automation cooperation for a steering task.

In the current study, the mutual adaptation behavior between driver grip strength and system authority was not observed, and thus further investigation will be addressed in a future study. Additionally, more participants including female drivers should be recruited in the future, since the current sample group was biased towards male drivers.

## Acknowledgment

The authors would like to thank Pan Xiuxi of Tokyo Institute of Technology for his assistance in realizing network communication between the Myo armband and haptic guidance system in driving simulator.